\documentclass[pdflatex,sn-mathphys-num]{sn-jnl}

\usepackage{graphicx}%
\usepackage{multirow}%
\usepackage{amsmath,amssymb,amsfonts}%
\usepackage{amsthm}%
\usepackage{mathrsfs}%
\usepackage[title]{appendix}%
\usepackage{xcolor}%
\usepackage{textcomp}%
\usepackage{manyfoot}%
\usepackage{booktabs}%
\usepackage{algorithm}%
\usepackage{algorithmicx}%
\usepackage{algpseudocode}%
\usepackage{listings}%
\usepackage{tikz}
\usepackage{cancel}

\usepackage{array}
\usepackage{caption}
\usepackage{booktabs}
\usepackage{graphicx}
\usepackage{geometry}
\def\mean#1{\left\langle#1\right\rangle}

\newcommand{\ei}{\varepsilon^\mathrm{int}_i}

\newcommand{\ep}{\varepsilon^\mathrm{PSF}_i}

\newcommand{\rg}{R^2_\mathrm{gal}}
\newcommand{\rp}{R^2_\mathrm{PSF}}

\theoremstyle{thmstyleone}%
\theoremstyle{thmstyletwo}%

\theoremstyle{thmstylethree}%

\begin{document}

\title{An ultra-high-resolution map of (dark) matter}


\author[1]{\fnm{Diana} \sur{Scognamiglio}}
\equalcont{These authors contributed equally to this work.}

\author[2]{\fnm{Gavin} \sur{Leroy}}
\equalcont{These authors contributed equally to this work.}

\author[3]{\fnm{David} \sur{Harvey}}
\equalcont{These authors contributed equally to this work.}

\author[2]{\fnm{Richard} \sur{Massey}}
\author[1]{\fnm{Jason} \sur{Rhodes}}
\author[4]{\fnm{Hollis B.} \sur{Akins}}
\author[5, 6]{\fnm{Malte} \sur{Brinch}}
\author[7]{\fnm{Edward} \sur{Berman}}
\author[8, 9, 10]{\fnm{Caitlin M.} \sur{Casey}}
\author[11]{\fnm{Nicole E.} \sur{Drakos}}
\author[12]{\fnm{Andreas L.} \sur{Faisst}}
\author[13, 9]{\fnm{Maximilien} \sur{Franco}}
\author[2]{\fnm{Leo W. H.} \sur{Fung}}
\author[14, 15]{\fnm{Ghassem} \sur{Gozaliasl}}
\author[16, 2]{\fnm{Qiuhan} \sur{He}}
\author[17]{\fnm{Hossein} \sur{Hatamnia}}
\author[1]{\fnm{Eric} \sur{Huff}}
\author[18, 19, 20]{\fnm{Natalie B.} \sur{Hogg}}
\author[21]{\fnm{Olivier} \sur{Ilbert}}
\author[22]{\fnm{Jeyhan S.} \sur{Kartaltepe}}
\author[23]{\fnm{Anton M.} \sur{Koekemoer}}
\author[10, 24]{\fnm{Shouwen} \sur{Jin}}
\author[25]{\fnm{Erini} \sur{Lambrides}}
\author[26]{\fnm{Alexie} \sur{Leauthaud}}
\author[2]{\fnm{Zane D.} \sur{Lentz}}
\author[27]{\fnm{Daizhong} \sur{Liu}}
\author[29, 2, 28]{\fnm{Guillaume} \sur{Mahler}}
\author[30]{\fnm{Claudia} \sur{Maraston}}
\author[8]{\fnm{Crystal L.} \sur{Martin}}
\author[7]{\fnm{Jacqueline} \sur{McCleary}}
\author[31]{\fnm{James} \sur{Nightingale}}
\author[17]{\fnm{Bahram} \sur{Mobasher}}
\author[32, 33]{\fnm{Louise} \sur{Paquereau}}
\author[13]{\fnm{Sandrine} \sur{Pires}}
\author[26]{\fnm{Brant E.} \sur{Robertson}}

\author[34]{\fnm{David B.} \sur{Sanders}}

\author[35]{\fnm{Claudia} \sur{Scarlata}}
\author[10, 36]{\fnm{Marko} \sur{Shuntov}}
\author[37, 38, 39]{\fnm{Greta} \sur{Toni}}
\author[2]{\fnm{Maximilian} \sur{von Wietersheim-Kramsta}}
\author[40, 41]{\fnm{John R.} \sur{Weaver}}

\affil[1]{Jet Propulsion Laboratory, California Institute of Technology, 4800 Oak Grove Drive, Pasadena, CA 91109, USA}
\affil[2]{Institute for Computational Cosmology, Department of Physics, Durham University, South Road, Durham DH1 3LE, United Kingdom}
\affil[3]{Laboratoire d’Astrophysique, École Polytechnique Fédérale de Lausanne (EPFL), Observatoire de Sauverny, CH-1290 Versoix, Switzerland}
\affil[4]{Department of Astronomy, The University of Texas at Austin, 2515 Speedway Blvd Stop C1400, Austin, TX 78712, USA}
\affil[5]{Instituto de Física y Astronomía, Facultad de Ciencias, Gran Bretaña 1111 Playa Ancha, Valparaíso, Chile}
\affil[6]{ Millennium Nucleus for Galaxies (MINGAL), Valparaíso, Chile }
\affil[7]{Department of Physics, Northeastern University, 110 Forsyth St. Boston, MA 02115, USA}
\affil[8]{Department of Physics, University of California Santa Barbara, Santa Barbara, CA 93106, USA}
\affil[9]{The University of Texas at Austin, 2515 Speedway Blvd Stop C1400, Austin, TX 78712, USA}
\affil[10]{Cosmic Dawn Center (DAWN), Denmark}
\affil[11]{Department of Physics and Astronomy, University of Hawaii, Hilo, 200 W Kawili St, Hilo, HI 96720, USA }
\affil[12]{Caltech/IPAC, 1200 E. California Blvd., Pasadena, CA 91125, USA}
\affil[13]{Université Paris-Saclay, Université Paris Cité, CEA, CNRS, AIM, 91191 Gif-sur-Yvette, France}

\affil[14]{Department of Computer Science, Aalto University, PO Box 15400, Espoo, FI-00 076, Finland}
\affil[15]{Department of Physics, Faculty of Science, University of Helsinki, 00014 Helsinki, Finland}
\affil[16]{Kapteyn Astronomical Institute, University of Groningen, PO Box 800, NL9700 AV Groningen, The Netherlands}
\affil[17]{Department of Physics and Astronomy, University of California, 900 University Ave., Riverside, CA 92521, USA}
\affil[18]{ Institute of Astronomy, University of Cambridge, Madingley Road, Cambridge, CB3 0HA, UK}

\affil[19]{Kavli Institute for Cosmology, University of Cambridge, Madingley Road, Cambridge, CB3 0HA, UK}

\affil[20]{Laboratoire univers et particules de Montpellier, CNRS \& Université de Montpellier, Parvis Alexander Grothendieck, Montpellier, France 34090}

\affil[21]{Aix Marseille Univ., CNRS, CNES, LAM, Marseille, France}
\affil[22]{Laboratory for Multiwavelength Astrophysics, School of Physics and Astronomy, Rochester Institute of Technology, 84 Lomb Memorial Drive, Rochester, NY 14623, USA}
\affil[23]{Space Telescope Science Institute, 3700 San Martin Drive, Baltimore, MD 21218, USA}
\affil[24]{DTU Space, Technical University of Denmark, Elektrovej 327, 2800 Kgs. Lyngby, Denmark}

\affil[25]{NASA Goddard Space Flight Center, Code 662, Greenbelt, MD 20771, USA}
\affil[26]{Department of Astronomy and Astrophysics, University of California, Santa Cruz, 1156 High Street, Santa Cruz, CA 95064, USA}
\affil[27]{Purple Mountain Observatory, Chinese Academy of Sciences, 10 Yuanhua Road, Nanjing 210023, China}
\affil[28]{Centre for Extragalactic Astronomy, Durham University, South Road, Durham DH1 3LE, UK}
\affil[29]{STAR Institute, Quartier Agora - Allée du six Août, 19c B-4000 Liège, Belgium}

\affil[30]{Institute of Cosmology \& Gravitation, University of Portsmouth, Portsmouth, PO1 3FX, UK}
\affil[31]{School of Mathematics, Statistics and Physics, Newcastle University, Herschel Building, Newcastle-upon-Tyne, NE1 7RU, UK}
\affil[32]{Department of Space, Earth and Environment, Chalmers University of Technology, SE-412 96 Gothenburg, Sweden}
\affil[33]{Institut d’Astrophysique de Paris, UMR}
\affil[34]{Institute for Astronomy, University of Hawaii, 2680 Woodlawn Drive, Honolulu, HI 96822, USA}
\affil[35]{Minnesota Institute for Astrophysics, University of Minnesota, 116 Church Street SE, Minneapolis, MN 55455, USA}
\affil[36]{Niels Bohr Institute, University of Copenhagen, Jagtvej 128, 2200 Copenhagen, Denmark}
\affil[37]{Dipartimento di Fisica e Astronomia ``A. Righi'', Alma Mater Studiorum Università di Bologna, via Gobetti 93/2, 40129 Bologna, Italy}
\affil[38]{INAF – Osservatorio di Astrofisica e Scienza dello Spazio di Bologna, via Gobetti 93/3, 40129 Bologna, Italy}

\affil[39]{Zentrum f\"{u}r Astronomie, Universit\"{a}t Heidelberg, Philosophenweg 12, D-69120, Heidelberg, Germany}

\affil[40]{MIT Kavli Institute for Astrophysics and Space Research, 70 Vassar Street, Cambridge, MA 02139, USA}

\affil[41]{Department of Astronomy, University of Massachusetts, Amherst, MA 01003, USA}

\abstract{Ordinary matter—including particles such as protons and neutrons—accounts for only about one sixth of all matter in the Universe. The rest is dark matter, which does not emit or absorb light but plays a fundamental role in galaxy and structure evolution. Because it interacts only through gravity, one of the most direct probes is weak gravitational lensing: the deflection of light from distant galaxies by intervening mass. Here we present an extremely detailed, wide-area weak-lensing mass map, covering $0.77^\circ \times 0.70^\circ$, using high-resolution imaging from the \textit{James Webb} Space Telescope (JWST) as part of the COSMOS-Web survey. By measuring the shapes of 129 galaxies per square arcminute—many independently in the F115W and F150W bands—we achieve an angular resolution of $1.00 \pm 0.01'$. Our map has more than twice the resolution of earlier \textit{Hubble} Space Telescope maps, revealing how dark and luminous matter co-evolve across filaments, clusters, and under-densities. It traces mass features out to $z\sim2$, including the most distant structure at $z\sim1.1$. The sensitivity to high-redshift lensing constrains galaxy environments at the peak of cosmic star formation and sets a high-resolution benchmark for testing theories about the nature of dark matter and the formation of large-scale cosmic structure.}

\maketitle

\section*{Main}\label{sec1}
Dark matter forms the gravitational backbone of the Universe, shaping the formation of galaxies and the large-scale structure of the cosmos. Yet, its distribution, particularly on intermediate and small scales, has remained difficult to map. Weak gravitational lensing, the slight but coherent distortion of background galaxy shapes by foreground mass, offers a direct way to probe this elusive component, without relying on assumptions about its physical state.
However, weak lensing maps so far have been limited in either resolution or sensitivity, obscuring the fine-grained structures that underpin the cosmic web \cite{Massey2007, Oguri2017, Martinet2018}.\\

\begin{figure*}[h!]
    \centering  \includegraphics[width=0.75\textwidth]{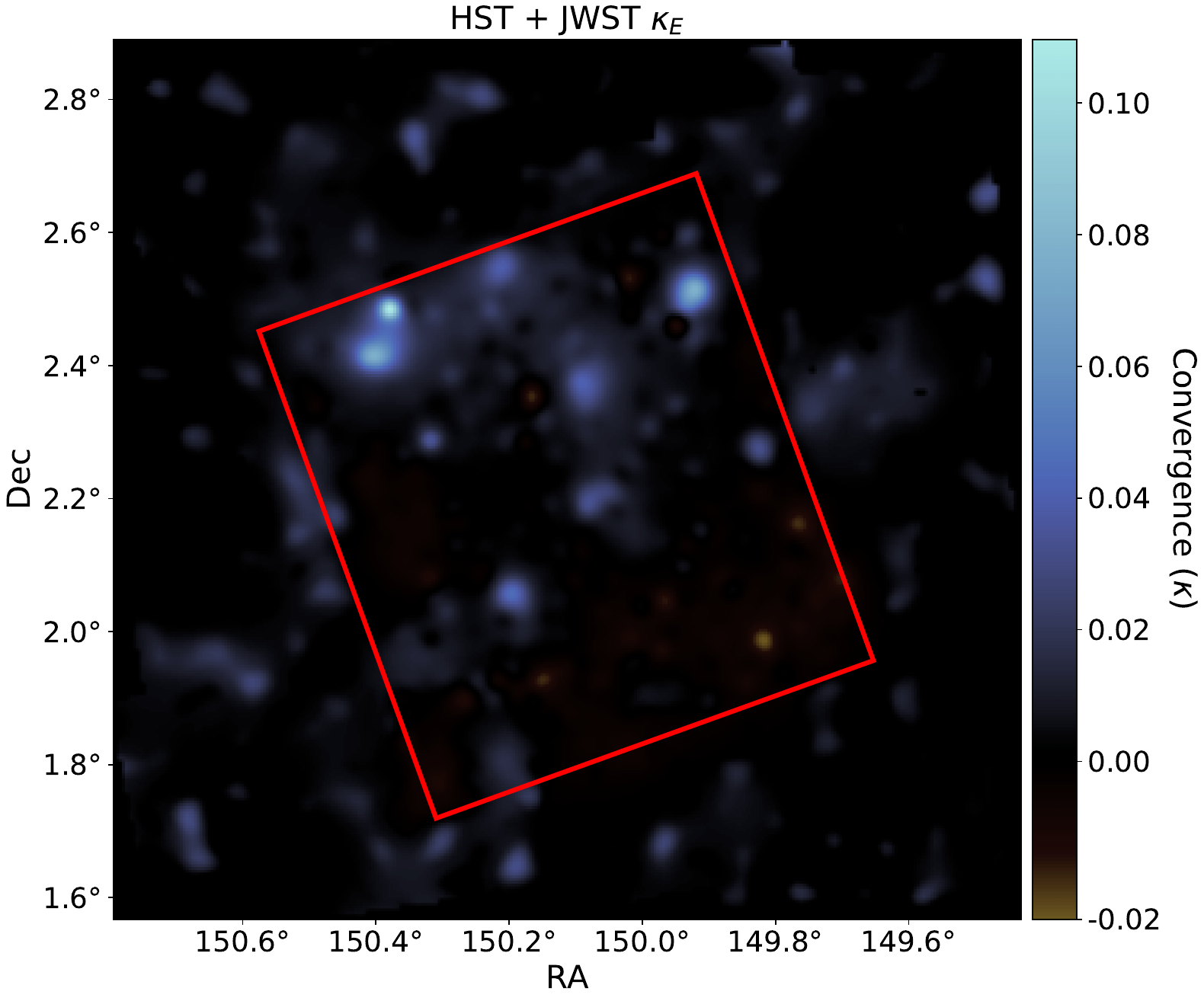} 
    \caption{Map of cosmic structure, from measurements of weak gravitational lensing with JWST and HST. Colours show the $E$-mode convergence ($\kappa$), where positive (blue) regions indicate mass overdensities and negative (brown) regions represent underdensities along the line of sight. The central region, outlined in red, corresponds to the COSMOS-Web footprint observed with JWST at a spatial resolution of $\sigma=1.0$~arcmin. It is embedded within a wider HST-based map at $\sigma=2.4$~arcmin resolution, to reduce edge effects and highlight the enhanced resolving power of JWST.}  
    \label{fig:k_maps_F150_F115}
\end{figure*}

\noindent
Here, we present a high-resolution mass map over a wide area of sky, derived from weak lensing measurements in the COSMOS-Web field using the \textit{James Webb} Space Telescope (JWST). By resolving the shapes of 129 galaxies per square arcminute---nearly double the number achieved with the \textit{Hubble} Space Telescope (HST)---we trace dark matter structures with a spatial resolution of 1 arcmin (Fig.~\ref{fig:k_maps_F150_F115}). This level of detail enables us to detect not only massive galaxy clusters, but also filamentary structures and underdense regions that were previously inaccessible.\\

\noindent
We focus on the COSMOS field---one of the most intensively studied regions of the sky---large enough to encompass a representative sample of large-scale structure, and covered by deep, high-resolution imaging and spectroscopy from the ultraviolet to the mid-infrared \cite{Scoville_2007}. Located on the celestial equator, it is accessible to observatories in both hemispheres.
Decades of observations from nearly all major ground- and space-based telescopes have provided a comprehensive multiwavelength view of this field \cite{Finoguenov_2007, Scoville_2007, Massey2007_Nature, Laigle2016, Smolcic2017, Liu_2019, Casey_2023}. These efforts have been significantly extended by 255 hours of JWST Near-Infrared Camera (NIRCam) imaging from the COSMOS-Web survey \cite{Casey_2023}, which covers a contiguous area of 0.54 square degrees in four bandpasses (F115W, F150W, F277W, and F444W).
In this study, we use imaging in the F115W and F150W bands, centred at 1.1~$\mu$m and 1.5~$\mu$m, respectively, with a spatial resolution of 0.02~arcseconds per pixel \cite{franco2025, shuntov2025, harish2025}, to perform shape measurements.\\

\noindent
Gravitational lensing affects the path of light from distant galaxies as it travels through the gravitational potential of any mass along the line of sight, including both ordinary baryonic matter (stars, gas, and dust) and dark matter \cite{Kaiser1993, Seitz1995}. In the weak gravitational lensing regime, this deflection induces a shear distortion of only a few per cent in the apparent shapes of background galaxies \cite{BertSchn01}. This distortion is roughly ten times smaller than the typical intrinsic ellipticity of galaxies, which is shaped by features such as spiral arms or bars. (We ignore `intrinsic alignments' between galaxies that add coherent systematics on smaller scales than probed here, but which should be considered in future applications of the shear catalogue.)
The resolution of a weak lensing map is therefore the area required to enclose about 100 resolved background galaxies, so that the shear signal rises above the shape noise. This number reflects a trade-off between spatial resolution and statistical uncertainty.
\\

\begin{figure*}[h!]
    \centering  \includegraphics[width=0.85\textwidth]{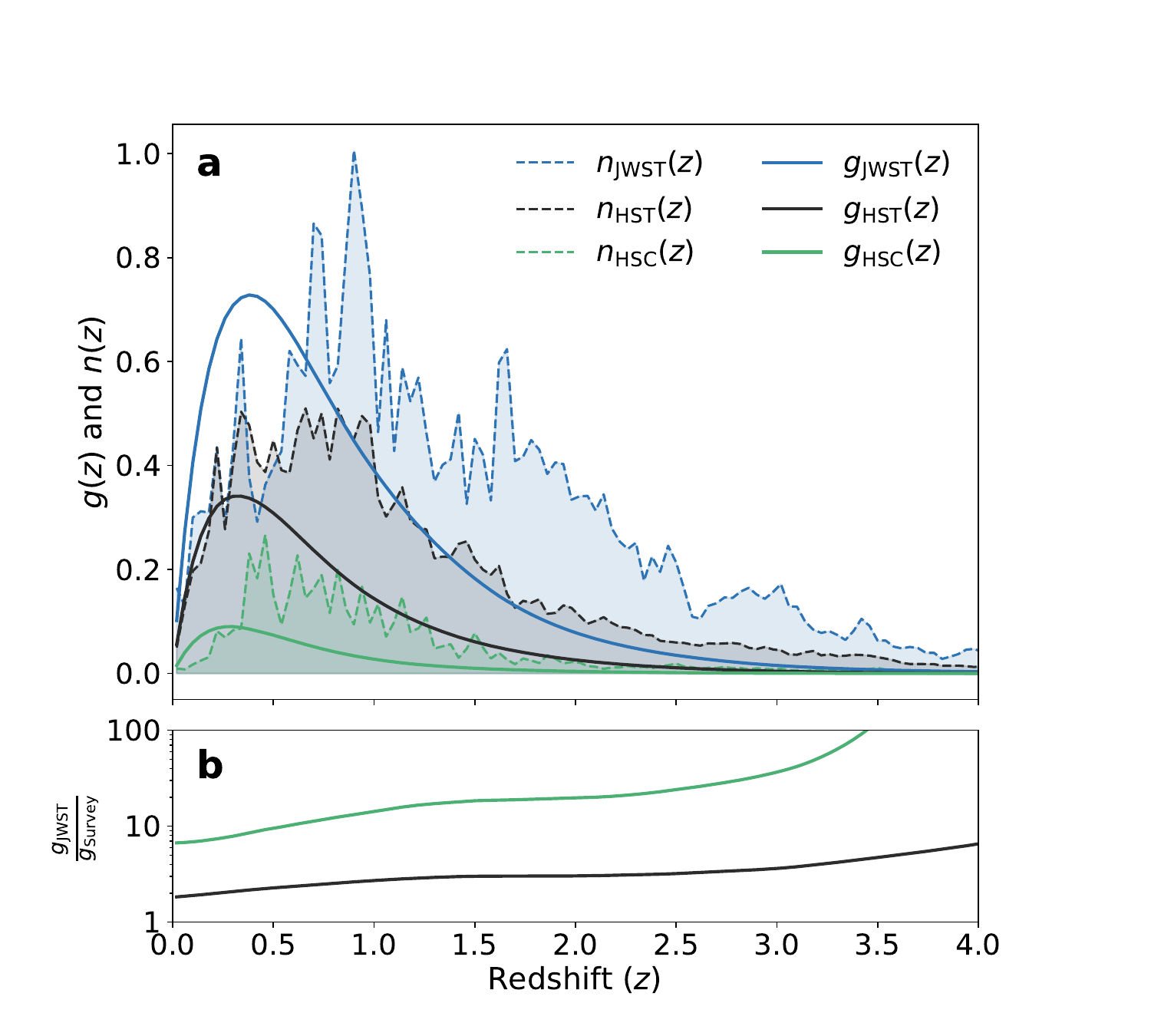} 
    \caption{Sensitivity of weak lensing to mass at different cosmic distances, quantified by redshift, $z$. \textbf{a}, Dashed lines show the measured number density, $n(z)$ of galaxies whose shapes, distorted by lensing, are resolved by JWST, HST, and HSC, in bins of width $\Delta z = 0.04$. JWST resolves the shapes of progressively more distant galaxies than other telescopes. Solid lines show the implied sensitivity functions $g(z)$ to foreground mass. \textbf{b}, Ratio of the JWST lensing sensitivity to that of HSC and HST. JWST has higher sensitivity at all redshifts, but especially at high redshift.
    }
    \label{fig:sensitivity}
\end{figure*}

\noindent
The sensitivity of a weak lensing map also depends on geometry. Like a magnifying glass that works best when held midway between the eye and the object, gravitational lensing is most sensitive to mass located half-way between the observer and the source galaxies. This behaviour is described by the lensing efficiency function $g(z)$, which peaks at lower redshifts (distances) than the distance to typical background objects, described by the source redshift distribution $n(z)$ (Fig.~\ref{fig:sensitivity}a).  
The lensing efficiency function for JWST peaks at $z=0.38$, for HST at $z=0.34$, and for the Subaru Telescope's Hyper-Suprime Camera (HSC) at $z=0.30$. 
Unlike most observables, for a fixed angular scale on the sky, weak lensing is insensitive to mass in the very nearby Universe because the lensing efficiency drops to zero as the lens approaches the observer.\\

\noindent
Previous weak lensing maps were limited in sensitivity, resolution, and area. From the ground, even the best telescopes$-$such as those used in the HSC, Kilo Degree Survey, and Dark Energy Survey$-$must contend with the blurring effects of Earth's atmosphere, and typically resolve the shapes of only 7–19 galaxies per square arcminute \cite{Hamana2020,des_maps,kids_maps,Jarvis2016}. These surveys cover a large area of sky, but are limited by coarse angular resolution and are sensitive primarily to structures at low redshift. As a result, only the most massive and extended structures, such as rare superclusters with masses up to 10$^{15} M_{\odot}$, appear prominently in ground-based lensing maps. The HST, which observes above the atmosphere, resolves the shapes of $\sim$71 galaxies per square arcminute \cite{Massey2007_Nature, Schrabback2010, Amara2012}, enabling maps with sufficient resolution of $\sim$\,2.4~arcmin to begin revealing clusters and filamentary features in the cosmic web (using a different definition of resolution, \cite{Massey2007_Nature} reported 1.2~arcmin resolution; see Methods Section~\nameref{sec:massrecon}).\\

\noindent
JWST’s deep, high-resolution imaging enables the measurement of weak lensing from a much larger number of galaxies, and at higher redshifts than previously possible. This improvement is reflected in the relative lensing sensitivity, where JWST outperforms both HSC and HST, particularly at high redshift (Fig.~\ref{fig:sensitivity}b). We measure shear for 108 galaxies per square arcminute in each of F115W and F150W. The final catalogue is constructed from the unique sources in each band, together with those in common for which the two shear estimates are averaged per galaxy, giving an effective density of 129 galaxies per square arcminute. The availability of repeat measurements in two bands helps reduce photon-counting (shot) noise when estimating shear, although the dominant source of uncertainty remains intrinsic shape noise.
Multi-wavelength ancillary data provide photometric redshifts for all galaxies (for more details see Methods Section~\nameref{catalogs}). We adopt photometric redshifts from \textsc{LePhare}, defined as the median of the likelihood distribution \cite{Ilbert_2006, 2011Arnouts}. For the sample of galaxies with measured shapes, this yields a median redshift of $z_{\mathrm{med}} \simeq 1.15$ \cite{shuntov2025}.\\

\begin{figure}[h!]
      \centering  
\includegraphics[width=0.75\columnwidth]{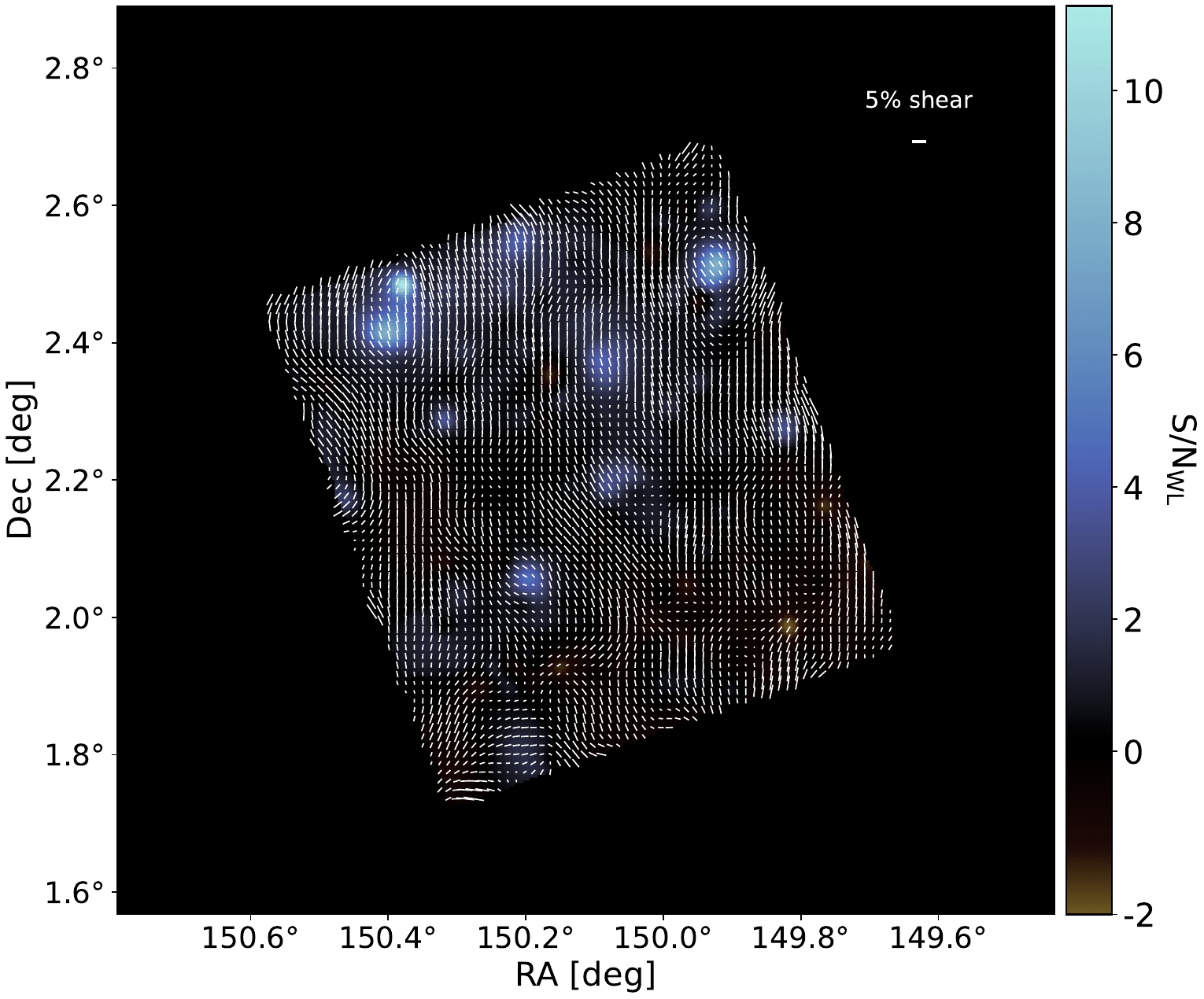}\caption{Measured shear distortion of distant galaxies.
The shapes of all galaxies from F115W and F150W-band imaging are averaged in $1.1'\times 1.1'$ pixels, followed by denoising through smoothing with a Gaussian filter of width $\sigma=1.2$~arcmin. The white line segments depict the local amplitude and orientation of the shear. The background image shows the signal-to-noise ratio, S/N$_{\mathrm{WL}}$, of the amplitude of the ``curl-free $E$-mode'' component of this pseudo-vector field. The reference bar in the top-right corner corresponds to $\gamma=0.05$ (5\% distortion relative to unlensed shapes).}
      \label{fig:wisker_maps}
\end{figure}

\noindent
Leveraging this unprecedented dataset, we construct the a detailed map of mass in large-scale structure to date (Fig.~1). 
Brighter regions indicate lines of sight with higher lensing convergence, which is proportional to the density of dark and luminous matter $\rho(x,y,z)$, multiplied by sensitivity function $g(z)$ and projected on the sky. The JWST shear measurements provide a spatial resolution of $1.00\pm0.01$~arcminute, more than a factor of two improvement over the previous large-scale COSMOS weak lensing map obtained with HST \cite{Massey2007_Nature}. We use a multiscale filtering technique \cite{Starck_2006} that identifies structures at different scales, which are defined as spatial scales in angular units (arcminutes), corresponding to physical sizes on the sky. This technique allows us to simultaneously identify both small features, such as halos around galaxy groups at low mass ($\sim$$10^{13}$–$10^{14}\,M_\odot$) or high redshift ($z \sim 1.1$), as well as the filamentary structures predicted to connect them \cite{Bond1996}.\\

\noindent
The enhanced sensitivity to lensing by mass at high redshift ($z > 1$) improves our ability to detect more distant structures, extending the reach of mass mapping beyond previous observational limits. This advancement enables direct constraints on the growth of structure, the assembly of dark matter halos, and the role of environment in shaping galaxy evolution during the peak epoch of star formation. The same sensitivity has already revealed over 100 strong lenses in the COSMOS-Web field \cite{Nightingale_2025, mahler2025, hogg2025}, highlighting JWST’s exceptional lensing capabilities and opening new avenues for joint weak and strong lensing studies of cosmic shear.\\

\noindent
The mass reconstruction is based on the observed weak lensing shear field, which captures the coherent distortions induced by foreground mass on background galaxy shapes. Pronounced shear alignments (Fig.~\ref{fig:wisker_maps}) in regions of high signal-to-noise, S/N$_{\mathrm{WL}}$ are consistent with lensing by massive structures at intermediate redshifts ($z \sim 0.2 - 1.1$) and high source density. These features coincide with convergence peaks and trace projected mass overdensities associated with galaxy groups and clusters. To enhance the signal, the shear field is binned into $1.1' \times 1.1'$ pixels and smoothed with a Gaussian of width $\sigma=1.2'$, highlighting its alignment with the underlying large-scale structure and reinforcing JWST’s ability to probe the gravitational imprint of cosmic structure.\\

\begin{figure}[h!]
      \centering  
\includegraphics[width=\textwidth]{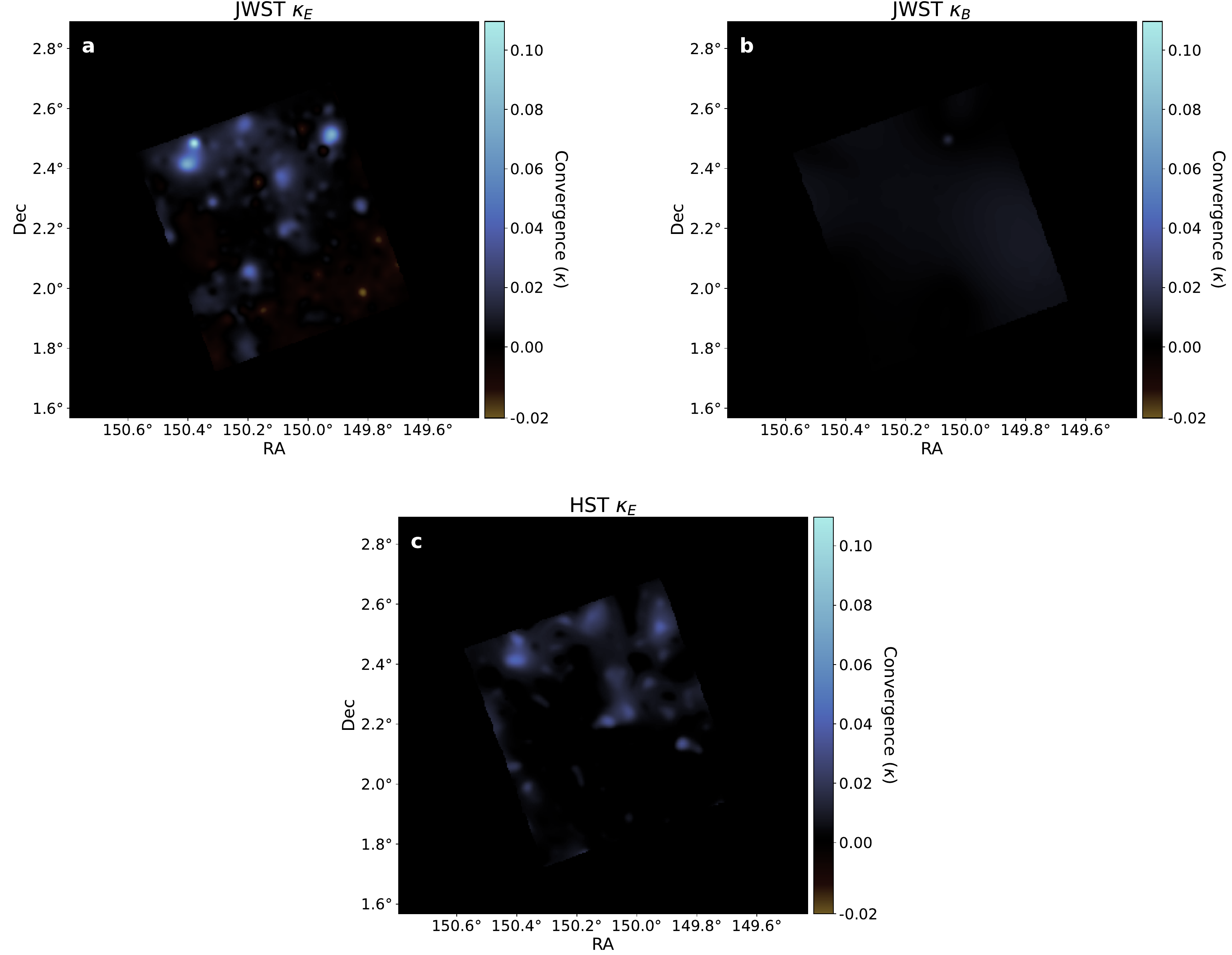}
\caption{Maps of the weak gravitational lensing  convergence from JWST and HST. \textbf{a}, JWST-based $E$-mode convergence map, showing the projected distribution of mass. \textbf{b}, JWST $B$-mode (null test) convergence map, which could contain systematic biases but is more than an order of magnitude lower in amplitude. \textbf{c}, HST-based $E$-mode convergence map, cropped to the JWST COSMOS-Web footprint, illustrating the same field at lower resolution \cite{Massey2007_Nature}. In all panels, the colour scale is identical, with brighter tones indicating mass overdensities, and an in-painting technique is applied to handle masked regions.}
    \label{fig:ebmode_F150}
\end{figure}

\noindent
To assess the level of residual systematics in mass maps, we note that the reconstructed map is complex-valued. The real part (Fig.~\ref{fig:ebmode_F150}a, c) corresponds to the so-called ``$E$-mode'' signal, which traces the curl-free component of shear and is produced by the projected foreground mass distribution \cite{Schneider_2002}. As expected, the HST-based $\kappa_{\mathrm{E}}$ map (Fig.~\ref{fig:ebmode_F150}c) reveals large-scale structure, while the JWST-based map (Fig.~\ref{fig:ebmode_F150}a) shows significantly finer substructures, enabled by higher spatial resolution and deeper imaging.
The imaginary part of the shear field represents the divergence-free ``$B$-mode'', which to first order is expected to be consistent with zero in standard cosmology. Since $E$- and $B$-modes are equally affected by statistical noise and equally likely to be affected by imperfect point spread function (PSF) correction or mask effects, the $B$-mode is useful as a diagnostic of potential systematics in the data. 
The measured $B$-mode amplitude is more than an order of magnitude lower than the $E$-mode (Fig.~\ref{fig:ebmode_F150}b), confirming that potential systematic biases in our analysis have been well controlled.\\

\begin{figure}[h!]
      \centering  
\includegraphics[width=\textwidth]{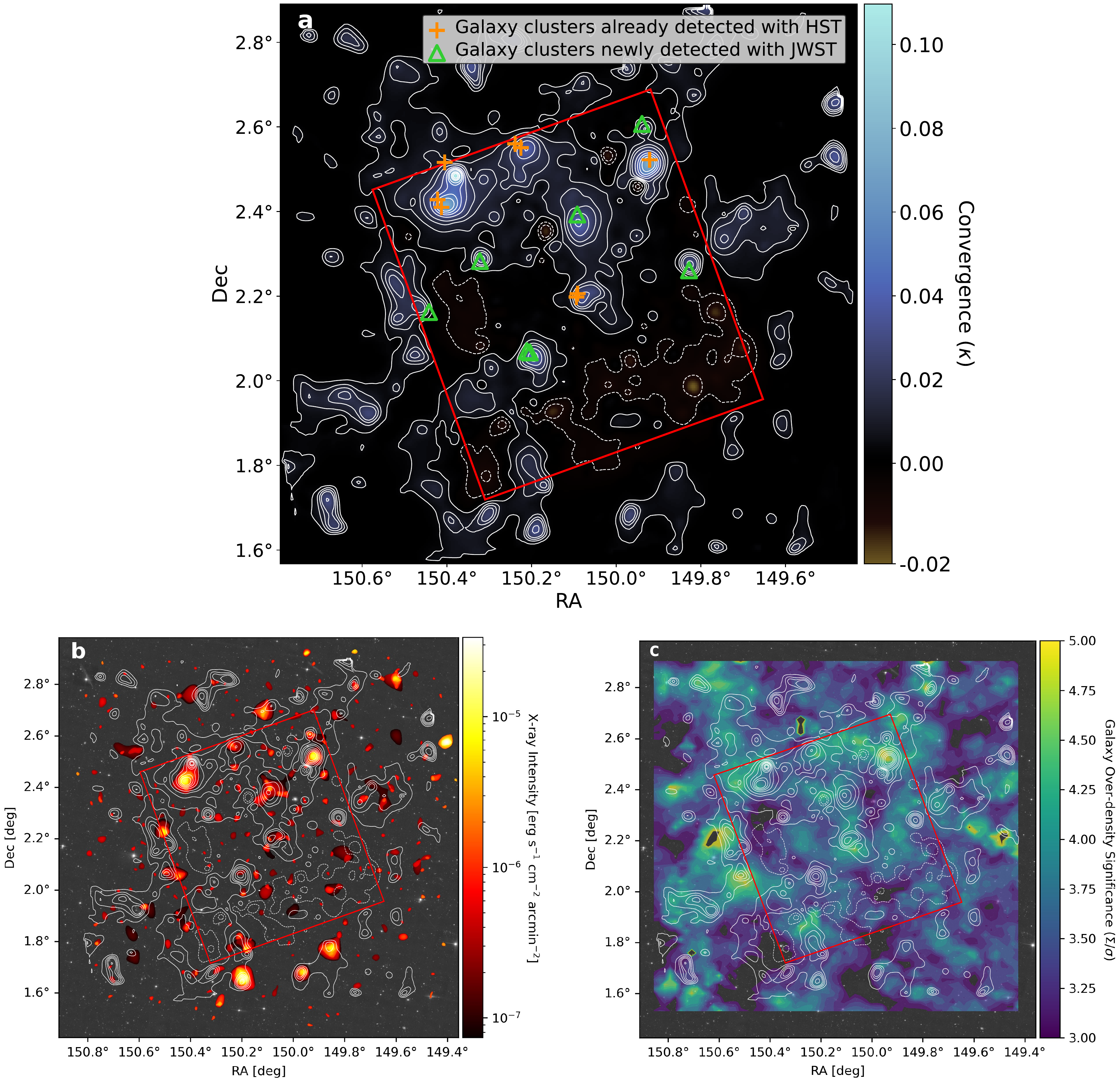}
\caption{Maps of weak gravitational lensing, X-ray emission, and galaxy density in the COSMOS field. \textbf{a–c}, Combined JWST (red square; F115W+F150W) and HST (surrounding area) weak lensing convergence map ($\kappa$), with contours at $\kappa=0.004$ and uniformly spaced by $\Delta\kappa=0.005$, matching the levels used in \cite{Massey2007_Nature} for direct comparison. Dashed sections mark the negative portions of the contours, corresponding to the same first convergence levels. \textbf{a}, Green triangles represent galaxy clusters newly detected in the JWST mass map, while orange crosses indicate clusters detected in both JWST and HST maps and matching known systems from XMM-\textit{Newton} and \textit{Chandra}, whose mass and redshift mean they should be (and are) detected at S/N$_\mathrm{det}>3$ in weak lensing. \textbf{b}, X-ray emission from hot gas in massive halos, traced with data from XMM-\textit{Newton} and \textit{Chandra}, with overlaid white $\kappa$ contours. \textbf{c}, Galaxy overdensity significance map ($\Sigma/\sigma$), weighted by the weak lensing sensitivity function $g(z)$, tracing the projected distribution of luminous matter, with white $\kappa$ contours marking the total mass.}

    \label{fig:mass_multi_panel}
\end{figure}

\noindent
The gravitational lensing convergence map derived from JWST data (Fig.~\ref{fig:mass_multi_panel}a) reveals the projected distribution of total matter---both dark and baryonic---across the COSMOS field. The increased resolution and source density afforded by JWST enable precise mapping of mass structures across a wide range of spatial  scales and redshifts. Notably, 15 known galaxy clusters previously detected via extended X-ray emission with XMM-\textit{Newton} and \textit{Chandra} \cite{Finoguenov_2007, Gozaliasl2019} are all recovered at a detection signal-to-noise SNR$_{\mathrm{det}} > 3$, considering a matching radius smaller than 2 arcmin. This constitutes a substantial improvement over earlier HST-based reconstructions, which detected only eight of these clusters, and validates the fidelity of our weak lensing measurements (see also Supplementary Table~\ref{tab:new_cluster}).\\

\noindent
Beyond isolated peaks, the JWST-based convergence map unveils a network of extended, low-amplitude features (Fig.~\ref{fig:mass_multi_panel}, dashed contours), that bridge cluster-scale overdensities. These structures probably trace dark matter filaments of the cosmic web, too diffuse to emit significant X-ray radiation or host large galaxy overdensities. Their detection is consistent with predictions from gravitational collapse in $\Lambda$CDM cosmology, and reflects JWST's enhanced sensitivity to the diffuse components of the matter field previously unresolved by HST \cite{Massey2007_Nature}.\\

\noindent
Comparison with X-ray and galaxy overdensity significance maps (Fig.~\ref{fig:mass_multi_panel}b, c) illustrates the coupling between dark matter, hot gas, and luminous galaxies. The X-ray map traces thermal emission from the hot intracluster medium in group- and cluster-scale halos, while the galaxy overdensity map, derived from COSMOS-2020 photometry \cite{COSMOS2020} and weighted by the lensing sensitivity function $g(z)$, shows fluctuations in the projected number density of galaxies. These fluctuations are expressed in units of the local standard deviation ($\Sigma/\sigma$), where $\Sigma$ represents the projected galaxy density in each pixel, and $\sigma$ is the standard deviation of the galaxy density within a reference region. This ratio, $\Sigma/\sigma$, is used to quantify the local signal-to-noise ratio, helping to highlight statistically significant structures. Peaks in the weak lensing map align with prominent X-ray emission, including two massive clusters near (right ascension (RA), declination (Dec)) $\approx$ (150.41°, 2.41°) and (149.92°, 2.52°). In addition, elongated contours forming an L-shaped structure---from RA $\approx$ 150.4° to 149.9°, and Dec $\approx$ 2.6° to 2.1°---reveal filamentary bridges between clusters, offering a direct lensing-based view of the cosmic web.\\

\noindent
The enhanced depth and redshift reach of JWST also reveal previously undetected structures. 
For example, the mass peak at (149.94°, 2.60°) corresponds to X-ray emission from a low-mass group at $z = 0.342$ despite a paucity of luminous galaxies, and the peak at (RA, Dec) $\approx$ (150.32°, 2.28°) is associated with X-ray sources at $z > 0.9$.
These results demonstrate JWST’s unique ability to detect massive structures even at redshifts where X-ray emission is dimmed by cosmological distance, and where galaxy overdensities become less prominent. At redshifts $1 \lesssim z \lesssim 2$, the cosmic noon epoch of peak star formation \cite{Madau_2014}, JWST’s high lensing sensitivity (Fig.~\ref{fig:sensitivity}b) allows weak lensing peaks to continue tracing large-scale structure. This range is also where our photometric redshift catalogue retains a high density of well-resolved background sources. This opens a new window into the environment of galaxies forming at cosmic noon and their evolution under the influence of active galactic nucleus feedback, gas quenching, and dark matter assembly \cite{Wechsler_2018}.\\

\noindent
While some weak lensing peaks coincide with regions of high X-ray intensity and galaxy overdensity significance (Fig.~\ref{fig:mass_multi_panel}b,c), we also identify mass peaks without clear counterparts in the projected X-ray emission or galaxy distribution---for example, at (RA, Dec) $\approx$ (150.21°, 2.06°) and (150.32°, 2.28°). These features may arise from underluminous or dark-matter-dominated structures, projection effects from multiple aligned systems along the line of sight, or mass concentrations at redshifts not optimally captured by the galaxy density weighting. In particular, we have identified significant lensing peaks that may result from the alignment of two structures at (RA, Dec) $\approx$ (150.21°, 2.06°) at low redshift (0.186 and 0.370) enabling the detection of these very low-mass objects.
\\

\noindent
Compared to HST, JWST provides roughly a twofold improvement in angular resolution, source galaxy density, and the number of significant detections of lensing structures \cite{Massey2007_Nature}. At $z \gtrsim 0.7$, JWST’s lensing sensitivity exceeds that of HST by a factor of two, and that of leading ground-based surveys by an order of magnitude. While the revealed large-scale structures of dark and baryonic matter are aligned in a way consistent with theoretical expectations,  JWST reveals them with much greater detail, extending into low-density regions and higher redshift regimes. This map provides a detailed view of the dark matter scaffolding underlying galaxy formation, and serves as a foundation for future studies of cosmic structure, feedback, and the evolution of the matter field over time. Looking ahead, JWST’s enhanced sensitivity at $1 \lesssim z \lesssim 2$ (Fig.~\ref{fig:sensitivity}b) will allow tomographic reconstructions of the dark matter environments of galaxies during cosmic noon---the epoch of peak star formation. Such measurements, not attempted in the present work, would provide a direct link between large-scale structure and galaxy evolution, including the effects of active galactic nucleus feedback, gas quenching, and dark matter assembly.

\renewcommand{\figurename}{Supplementary Figure} 
\renewcommand{\tablename}{Supplementary Table} 
\setcounter{figure}{0}
\section*{Methods}
\subsection*{COSMOS-Web imaging and catalogs}
\label{catalogs}
The COSMOS-Web survey is a 270-hour JWST treasury programme (PID 1727; PIs: Kartaltepe \& Casey;  \cite{Casey_2023}), originally planned for 255 hours, covering 0.54~deg$^{2}$ of the COSMOS field \cite{Scoville_2007, Capak_2007} using the Near-Infrared Camera (NIRCam) and 0.19~deg$^{2}$ with the Mid-Infrared Imager (MIRI). The NIRCam imaging data span four filters: F115W, F150W, F277W, and F444W \cite{franco2023}. We use the F115W and F150W filters for weak lensing measurements owing to their superior angular resolution, higher throughput, and lower background noise compared to the longer-wavelength bands.\\

\noindent
Observations were carried out across three epochs (January 2023, April-May 2023, and January 2024), using a 4TIGHT dither pattern to ensure uniform coverage and depth, reaching AB magnitudes of 26.9–27.4 in each band. All exposures were calibrated and processed through the JWST Pipeline \cite{bushouse_2024} v1.14.1, which includes the resample step to combine the individual exposures into mosaics, following standard procedures \cite{Koekemoer2011}, with \texttt{pixfrac} = 1 and a pixel scale of 0.02 arcsec per pixel. These choices minimize aliasing of the native pixel scale, which is critical for weak lensing analysis \cite{Rhodes_2007}.\\

\noindent
To extend the weak lensing analysis to larger spatial scales and reduce border effects in the mass map, we embed JWST data inside the HST ACS COSMOS survey \cite{Massey2007_Nature}. HST F814W images covering an area of 1.637~deg$^{2}$ provide shape measurements for 71 galaxies per arcmin$^2$, with a median AB magnitude of 25.1. Shapes were measured using the Rhodes, Refregier \& Groth (hereafter, RRG \cite{Rhodes_2000}) method \cite{Leauthaud07}, which deconvolves the PSF from galaxy shapes and has been calibrated on HST simulations to recover shear with $<$ 6\% bias across a wide range of sizes and fluxes.\\

\noindent
Source detection was performed with \texttt{SExtractor} \cite{Bertin1996}, using a combined $\chi^{2}$ image constructed from all four NIRCam bands (F115W, F150W, F277W, F444W), leveraging the full wavelength coverage to enhance detection of faint sources. Source detection was performed in single-image mode, using the same image for both detection and photometric measurement. A threshold of 2.33$\sigma$ above the background was applied to optimize the trade-off between completeness and reliability. Multiwavelength photometry was extracted from PSF-matched images convolved to the F444W PSF, using Kron elliptical apertures fitted to each source. However, for weak lensing shape measurements, only the F115W and F150W images were used, due to their higher spatial resolution and lower background, ensuring optimal shape recovery.\\

\noindent
The COSMOS2025 catalogue (Shuntov et al., submitted to A\&A) combines COSMOS-Web imaging with ancillary multiwavelength data, providing photometry, photometric redshifts, physical properties, and morphological measurements for over 780,000 sources across the field. From this sample, we apply a series of quality cuts based on peak surface brightness (\texttt{MU\_MAX}), total flux (\texttt{MAG\_AUTO}), effective diameter ($d$), and detection signal-to-noise ratio (SNR$_{\mathrm{det}}$) to select galaxies with robust shape measurements (see Subsect.~\nameref{sec:Shape_meas}). This yields 209,322 galaxies in the F115W band and 209,893 in F150W, corresponding to a summed raw number density of 216 galaxies per square arcminute across 0.54 deg$^2$. From these, we construct a final catalogue of unique galaxies by retaining the shear estimate from single-band detections and averaging the two measurements for galaxies detected in both filters. This reduces scatter while preserving unbiased means, and improves noise characterization by mitigating bias during shear-map smoothing. The final sample comprises 170,386 galaxies detected in both bands, 38,936 detected only in F115W, and 42,135 only in F150W, yielding an effective number density of 129 galaxies per square arcminute.\\

\noindent
We use the \textsc{LePhare} photometric redshift estimator \citep{Ilbert_2006, 2011Arnouts}, which fits galaxy templates to multiwavelength photometry. The method generates a redshift probability distribution for each galaxy by calculating the likelihood for each template. The median of this likelihood distribution is then taken as the best estimate of the galaxy’s redshift. This approach also allows for the simultaneous derivation of physical parameters of galaxies. For stars and artefacts, the redshift is set to 0, as detailed by \cite{shuntov2025}.\\

\noindent
To trace the large-scale galaxy distribution, we use the COSMOS2020 catalogue \cite{COSMOS2020}, which offers consistent multiband photometry and photometric redshifts across the full COSMOS field, to construct a projected galaxy density map (Fig.~\ref{fig:mass_multi_panel}c).

\subsection*{Modelling the Point Spread Function} 
The apparent shapes of galaxies are distorted by gravitational lensing and further convolved with the telescope’s PSF, which smears their observed morphology and must be accurately modelled to obtain unbiased shear measurements. The PSF is characterized by its size, $\rp$, and shape expressed through its 
ellipticity components, $\ep$ for $i\in\{1,2\}$, all constructed from second-order moments with a circular Gaussian weight function with radius $\sigma=$1.5 pixels. We tested several PSF modelling approaches for JWST, including {ShOpt} \cite{berman2024}, \textsc{PSFEx} \cite{Bertin2011}, and \textsc{WebbPSF} \cite{2014Perrin}, and found that the latter yields the lowest residual systematics in the final shear measurements. \textsc{WebbPSF} enables us to model the PSF in individual exposures and combine them into a composite PSF matched to the full-depth images. This approach remains robust in regions with few bright stars and performs reliably near exposure boundaries, where the PSF varies discontinuously with position on the sky.\\

\noindent
To quantify the impact of PSF modelling errors on shear estimation, we analyse the difference throughout the survey between the measured ellipticity of unsaturated stars and that of PSF models, $\delta\varepsilon^{PSF} = \varepsilon_{\mathrm{measured}}^{\mathrm{PSF}} - \varepsilon_{\mathrm{model}}^{\mathrm{PSF}}$. In the F115W band, we use 5,620 stars, yielding a mean residual ellipticity of $\langle \delta\varepsilon^{\mathrm{PSF}} \rangle = (-0.0035,\ 0.0005)$ and root-mean-square (RMS) values of $(0.0158,\ 0.0151)$ for $(\delta\varepsilon_{1}^{\mathrm{PSF}},\ \delta\varepsilon_{2}^{\mathrm{PSF}})$. For F150W, 6,125 stars yield $\langle \delta\varepsilon^{\mathrm{PSF}} \rangle = (0.0005,\ -0.0001)$ with RMS values of $(0.0104,\ 0.0094)$. These residuals are interpreted as spurious shear and used to construct PSF-induced $E$-mode convergence maps, $\kappa^{\mathrm{PSF}}$, shown in Supplementary Figure~\ref{fig:psf_residual_kappa}. \\

\noindent
When using \textsc{WebbPSF}, the PSF-induced convergence maps exhibit low-amplitude, spatially uncorrelated structure in both bands, with RMS values of $\sigma_{\kappa^{\mathrm{PSF}}}= 1.1\times 10^{-3}$ for F115W and $0.8 \times 10^{-3}$ for F150W. These values are more than an order of magnitude smaller than the cosmological lensing signal (see Fig.~\ref{fig:k_maps_F150_F115}), confirming that PSF model errors are subdominant. On the other hand, we find empirical methods like ShOpt and \textsc{PSFEx} to be unstable. This is mainly because the stacking of multiple exposures introduces sharp discontinuities in the PSF model, resulting in occasional, localised catastrophic failures in (assumed smooth) PSF interpolation. Across the survey footprint, variance in the PSF-induced convergence maps is $2.5\times$ higher for \textsc{PSFEx} and $12.5\times$ higher for ShOpt.

\subsection*{Galaxy shape measurement and shear estimation}
\label{sec:Shape_meas}
We measure the shapes of galaxies using \textsc{pyRRG} \cite{Harvey2021, a2744_pyrrg}, which calculates Gaussian-weighted second- and fourth-order multipole moments, $J_{ij}$ and $J_{ijkl}$, for $i,j,k,l \in \{1,2\}$, all corrected for convolution with the PSF and pixel response. An estimator of galaxy size is $\rg \equiv J_{11} + J_{22}$. The galaxy's ellipticity is defined as 
\begin{equation}
\left( \varepsilon_{1}^{\mathrm{gal}}, \varepsilon_{2}^{\mathrm{gal}} \right)=\left( \frac{2J_{12}}{R^{2}_{\mathrm{gal}}},\frac{J_{11}-J_{22}}{R^{2}_{\mathrm{gal}}} \right),
\end{equation}
and estimator of the line-of-sight shear is 
\begin{equation}
\widehat{\gamma_i} \equiv \varepsilon_i^{\mathrm{gal}} / G, \label{eqn:shear_estimator}
\end{equation}
the shear susceptibility $G$ depends on fourth-order moments \cite{Rhodes_2000}. Shear susceptibility describes the response of an intrinsically elliptical galaxy to shear: $\varepsilon_i^{\mathrm{gal}} = \varepsilon_i^{\mathrm{int}} + G \gamma_i$, valid in expectation over a galaxy ensemble with $\left\langle \varepsilon_{i}^{\mathrm{int}} \right\rangle = 0$.\\

\noindent
Because the fourth-order shape moments in $G$ are inherently noisy and enter in the denominator of shear estimator \eqref{eqn:shear_estimator}, we use the mean value of $G$ fitted as a function of galaxies' detection signal-to-noise ratio SNR$_{\mathrm{det}}$, $G=A (1 - e^{{-B}\mathrm{SNR}_{\mathrm{det}}}) + C$, where parameters $A$, $B$, and $C$ are fitted independently for each filter (see Supplementary Figure~\ref{fig:shear_susceptibility}). We empirically determined this functional form, which captures the expected flattening of $G$ at high SNR$_\mathrm{det}$, where measurement noise becomes subdominant. Over the range $5 < \mathrm{SNR}_{\mathrm{det}} < 30$, we find that the mean susceptibility is $\langle G^\mathrm{F115W} \rangle = 1.156$ and $\langle G^\mathrm{F150W} \rangle = 1.120$.\\

\noindent
To reduce measurement biases from noise and PSF modelling uncertainties \cite{Leauthaud07, High07}, we apply galaxy selection criteria on size, magnitude, and $\mathrm{SNR}_{\mathrm{det}}$, tailored to each NIRCam filter. For F150W, we retain galaxies with \texttt{MAG\_AUTO} $<$ 28, effective diameter $d > 2.2$ pixels, and detection $\mathrm{SNR}_{\mathrm{det}} >$ 4.0. For F115W, we adopt similar cuts on \texttt{MAG\_AUTO} and $d > 2.0$ pixels, but apply a more relaxed $\mathrm{SNR}_{\mathrm{det}}$ threshold of zero to preserve fainter detections. These thresholds are less restrictive than those used in previous HST analyses \cite{Leauthaud07}, owing to the improved automatic catalogue cleaning implemented in \textsc{pyRRG}. The selection cuts were optimized to enhance the weak lensing signal while minimizing systematics, maximising the $\kappa_{E} / \sigma_{\kappa_{B}}$ ratio without diluting the signal by excessive galaxy exclusion. \\

\noindent
After applying these cuts, we obtain robust shear measurements for 209,322 galaxies in the F115W band and 209,893 in F150W, i.e.\ 108 galaxies arcmin$^{-2}$ in each band. 
The amplitude of measured shears \eqref{eqn:shear_estimator} is consistent between band, and consistent with those derived from HST imaging: with mean values $\langle \widehat{\gamma_1}^\mathrm{F115W} \rangle = 0.0116$ and $\langle \widehat{\gamma_2}^\mathrm{F115W} \rangle = -0.0073$, and $\langle \widehat{\gamma_1}^\mathrm{F150W} \rangle = 0.0132$ and $\langle \widehat{\gamma_2}^\mathrm{F150W} \rangle = -0.0052$. 
The RMS scatter of shear measurements, $\sigma_{\widehat{\gamma_i}}\equiv \langle(\widehat{\gamma_i})^2\rangle$, is also consistent: $\sigma_{\widehat{\gamma_1}}^{\mathrm{F115W}} = \sigma_{\widehat{\gamma_2}}^{\mathrm{F115W}} = 0.30$, $\sigma_{\widehat{\gamma_1}}^{\mathrm{F150W}} = \sigma_{\widehat{\gamma_2}}^{\mathrm{F150W}} = 0.31$, and $\sigma_{\widehat{\gamma_1}}^{\mathrm{HST}} = \sigma_{\widehat{\gamma_2}}^{\mathrm{HST}} = 0.32$ \cite{Leauthaud07}. \\

\noindent
Comparing shear measurements between the two JWST bands, we notice that $\langle \widehat{\gamma_i}^{\mathrm{F115W}}-\widehat{\gamma_i}^{\mathrm{F150W}}\rangle^2\ll\sigma_{\widehat{\gamma_i}}^2$, with ratios of 0.13 for $\gamma_{1}$ and 0.12 for $\gamma_{2}$, indicating that imaging in the two bands (necessarily) has independent shot noise, but that the intrinsic shapes of galaxies are not significantly different between bands. \\


\noindent
To assess the total leakage of PSF modelling, correction and other biases into our shear measurements, we average the final shear estimators $\langle \widehat{\gamma}_{i} \rangle$ in bins of PSF model ellipticity, $\varepsilon_i^\mathrm{PSF}$ (Supplementary Figure~\ref{fig:psf_measurement_biases}). To interpret these, note that equation~(26) of \cite{Massey_2013} can be rewritten as
\begin{equation}
\mean{\widehat{\gamma_i}} 
      \approx
      \left(1+\frac{\delta(\rp)}{\rg}\right)
      \left(\cancelto{0}{\mean{\frac{\ei}{G}}}+\cancelto{0}{\mean{\gamma_i}}~~~\right) 
      - \mean{\frac{\rp}{G\rg}} \left( \mean{\delta\ep} 
      + \mean{\frac{\delta(\rp)}{\rp}}\,\ep \right) ,
      \label{eqn:shear_residuals}
\end{equation}
where $\delta\rp$ and $\delta\ep$ are residual errors in the size and ellipticity of the PSF model (compared to the truth rather than noisy, measured values), and where we have assumed that $\delta\ep$ is a function of neither $\ep$ nor $\rp$ in order to separate uncorrelated terms in angle brackets. Both terms in the first bracket vanish under the assumption that neither intrinsic ellipticity nor true shear depends on PSF ellipticity. This leaves a linear relation $\mean{\widehat{\gamma_i}}=a_i+b_i\ep$,
whose free parameters we fit in Supplementary Figure~\ref{fig:psf_measurement_biases}.
Equating coefficients with equation~(\ref{eqn:shear_residuals}) and adopting average values of $\langle{\rg/\rp}\rangle\approx1.2$ and $\langle G \rangle$ as above, we estimate the residual PSF model ellipticity, averaged across all observations 
\begin{equation}
\mean{\delta\ep}\approx -1.2\langle G\rangle a_{i}=(0.0124,\ -0.0069) \mathrm{~for~F115W,~and~} (0.0069, -0.0055) \mathrm{~for~F150W}.
\end{equation}

\noindent
Assuming a typical PSF ellipticity of $\langle \varepsilon_i^\mathrm{PSF} \rangle \sim 0.015$, and interpreting equation~(26) of \cite{Massey_2013} as shear measurement biases \cite{Massey2007}, we estimate additive shear measurement bias $c_i = a_i + b_i \langle \varepsilon_i^\mathrm{PSF} \rangle=(-0.0011,-0.0098)$ for F115W, and $(-0.0152,-0.0074)$ for F150W. These additive shear systematics are mainly due to the smallest galaxies, but our choice of size cuts ensures they are subdominant to the (statistical noise on the) shear signal in a typical resolution element of the map.\\

\noindent
To quantify residual correlations in shear estimators as a function of angular separation $\theta$ (not just on average), we employ the $\rho_{n}(\theta)$ statistics \cite{Jarvis2016}, which quantify residual correlations in PSF-corrected shapes. 
With \textsc{WebbPSF}, we find that all $\rho_n$ functions remain below $10^{-5}$ at separations $\theta > 1'$ (Supplementary Figure~\ref{fig:psf_rho_statistics}), the scales to which weak lensing is sensitive. This indicates that residual PSF correlations are negligible and that additive shear systematics are well controlled, validating the use of \textsc{WebbPSF} for mass mapping with JWST.\\

\noindent
In principle, blending of light from galaxies on adjacent lines of sight can lead to multiplicative shear measurement biases, particularly in regions of high galaxy density. However, JWST's high imaging resolution 
substantially reduces the impact of blending (compared with HST, JWST's spatial resolution is a factor $\gtrsim$$2^2$ higher, and its pixels a factor $2.5^2$ finer, but the density of galaxies is only 1.8~times higher). In addition, the hot/cold source separation used in our analysis helps to mitigate artificial, spurious deblending. To test the impact of blending, we remove adjacent pairs of galaxies from our analysis: we find that even the most conservative cuts change the mean shear in pixels by only $\sim$2\%. Deep JWST imaging thus provides high-quality shear measurements even in dense environments such as galaxy clusters. Indeed, we found the dominant sources of uncertainty to be in modelling the JWST PSF rather than pixel-scale or astrophysical effects. Consequently, the most informative validation is obtained by comparing JWST shear estimates directly against stable and well-understood HST measurements of the same galaxies (see also \cite{a2744_pyrrg}).

\subsection*{Mass map reconstruction} 
\label{sec:massrecon}
The reconstruction of the matter distribution begins by binning the observed shear field into \textbf{ $15''\times15''$} pixels, each containing approximately eight JWST galaxies.
To convert the observed shear field into $E$- and $B$-mode convergence ($\kappa$), we apply an enhanced version of the KS+ algorithm \cite{Pires2009a, Pires_2020}. This approach iteratively accounts for the non-linear relationship between the reduced shear and the true shear, where the reduce shear is defined as $g\equiv \gamma/(1-\kappa)$ \cite{Schneider_2002}, and employs in-painting to recover the shear field across pixels containing no galaxies (for JWST data, 0.28\% of the area, due to bright stars or the survey footprint edge). Missing regions are reconstructed using a wavelet-based sparsity constraint on the power spectrum of the convergence map \cite{Pires_2020}. Notably, binning HST shear measurements in the same $15''\times15''$ pixels would leave a 11.4\% masked area, owing to more extensive scattered light contamination from stars. However, the brightest star in the region (RA, Dec = 150.23°, 2.63°) lies inside the HST survey footprint, but outside the JWST footprint.\\

\noindent
We filter the weak lensing convergence map using the well-tested and widely used MRLens algorithm \cite{Starck2006}. 
This uses the \textit{\`a trous} wavelet transform and the False Discovery Rate (FDR) method to filter simultaneously both large-scale and small-scale structures.
The FDR method provides a rigorous framework adaptive to the data to suppress noise at each scale and quantify uncertainties \cite{benjamini1995}.
In this process, some small scales are completely removed, while larger scales are thresholded. The choice of which scales to remove and the level of thresholding applied to each wavelet scale are determined by the FDR method, under the assumption that the noise follows a Gaussian distribution, in order to limit the fraction of spurious peaks to below 1\% at all scales.\\

\noindent
The wavelet scales are directly defined by the pixel size and increase in powers of two. So, the pixel size can be optimized to limit the scales to be removed. The resolution of the filtered map is thus defined by the smallest wavelet scale used to build the filtered convergence map. To optimize the resolution, we vary the pixel scale smoothly, allowing the MRLens algorithm to select which wavelet scales to include. In the final map (Fig.~\ref{fig:k_maps_F150_F115}) with a pixel size of 15$''$ the signal is suppressed in the first two wavelet scales, and we use the following thresholds: $4.63\sigma_{\kappa}$, $3.34\sigma_{\kappa}$ and $1.96\sigma_{\kappa}$ in the next three wavelet scales.\\


\noindent
Noise properties of the map vary across scales and are entangled with the local signal in a non-trivial manner. To assess the level of noise in the $\kappa_{\mathrm{E}}$ map, we use the $\kappa_{\mathrm{B}}$ map, which ideally contains no cosmological signal \cite{Schneider_2002}. Applying the same MRLens filtering to the $B$-mode shear component, with identical noise estimates and FDR thresholds for consistency, yields mean $B$-mode signal more than an order of magnitude smaller than the similarly-filtered $E$-mode signal. We use its standard deviation, $\sigma_{\kappa_{\mathrm{B}}}$ to define the weak lensing signal-to-noise ratio 
$\mathrm{SNR}_{\mathrm{WL}} = W^j_{\kappa_{\mathrm{E}}}/ (\sigma_{\kappa_{\mathrm{B}}} *\alpha_j)$, where $W^j_{\kappa_{\mathrm{E}}}$ corresponds to the lensing signal measured at wavelet scale $j$ and $\alpha_j$ is a normalization factor that depends on the shape of the wavelet kernel \cite{Starck2006}. This approach accounts for spatial variations in galaxy density and imaging depth, enabling accurate assessment of signal significance across the field. Averaged across the map, at the smallest (`$W^3_{\kappa_{\mathrm{E}}}$') wavelet scale, $\langle\mathrm{SNR}_{\mathrm{WL}}\rangle=1.9$ per resolution element. \\

\noindent
Alternative map reconstruction algorithms include diffusion-based methods \cite{aoyama2025} or other machine-learning techniques \cite{Cha_2025}. However, these remain dependent on large sets of realistic training simulations and may be sensitive to mismatches between simulated and observed data.\\

\noindent
To identify individual dark matter structures, we do not actually use the denoised MRLens convergence maps. Instead, we use a multiscale peak detection algorithm based on wavelet decomposition \cite{LEROY2023}. This method searches for local maxima (SNR$_{\mathrm{det}} > 3$) across consecutive wavelet scales, particularly those corresponding to structures of size of 1 to 4~arcmin. Compared to MRLens, this approach is able to isolate compact features associated with galaxy groups and clusters while suppressing large-scale modes.\\

\noindent
Several newly detected mass peaks in the JWST weak lensing map are spatially coincident with X-ray-detected clusters observed by XMM–\textit{Newton} and \textit{Chandra} \cite{Gozaliasl2019}, with an average positional offset of $0.78\pm 0.34$~arcminutes (Supplementary 
Table~\ref{tab:new_cluster}). This marks an improvement over the HST-based mass map, which yields a mean offset of $1.19\pm0.54$~arcminutes when comparing a subset of HST peaks selected with SNR$_{\mathrm{det}} > 3$ and matched within 2$'$. The JWST map includes 15 cluster matches, compared to only up to 8 from the HST-based reconstruction in the same area.

\section*{Data Availability}
The JWST data (Programme ID: General Observer 1727) are publicly available at \url{https://exchg.calet.org/cosmosweb-public/DR0.5/}.\\
The HST data (Programme IDs: General Observer 9822 and 10092) are publicly available at \url{http://irsa.ipac.caltech.edu/data/COSMOS/}.\\
The XMM-\textit{Newton} dataset (Programme ID: 020336) is publicly available in
staged releases through the IPAC/IRSA website \url{http://irsa.ipac.caltech.edu /data /COSMOS/}.\\
The \textit{Chandra} data (Programme IDs: 901037) are publicly available at \url{https://irsa.ipac.caltech.edu/data/COSMOS/gator_docs/cosmos_chandraxid_colDescriptions.html}.\\ 
The weak lensing mass maps will be made publicly available following the acceptance of this paper. 


\section*{Acknowledgments}
D.S. carried out this research at the Jet Propulsion Laboratory, California Institute of Technology, under a contract with the National Aeronautics and Space Administration (80NM0018D0004). Support for this work was provided by NASA grants JWST-GO-01727 and HST-AR15802 awarded by the Space Telescope Science Institute, operated by the Association of Universities for Research in Astronomy, Inc., under NASA contract NAS 5-26555. 
G.L., R.M. and, M.v.W.K. acknowledge support from STFC via grant ST/X001075/1, and the UK Space Agency via grant ST/W002612/1 and InnovateUK (grant no. TS/Y014693/1). D.H.\ was supported by the Swiss State Secretariat for Education, Research and Innovation (SERI) under contract number 521107294. This project has received funding from the European Union’s Horizon 2020 research and innovation programme under the Marie Skłodowska-Curie grant agreement No 101148925. French COSMOS team members are partly supported by the Centre National d’Etudes Spatiales (CNES). O.I. acknowledges the funding of the French Agence Nationale de la Recherche for the project iMAGE (grant ANR-22-CE31-0007). G.M. is supported in Durham by STFC via grant ST/X001075/1, and the UK Space Agency via grant ST/X001997/1. S.J. acknowledges the European Union’ Marie Skłodowska-Curie Actions grant No. 101060888, and the Villum Fonden research grants 37440 and 13160. N.E.D acknowledges support from NSF grants LEAPS-2532703 and AST-2510993. D.B.S. gratefully acknowledges support from NSF Grant 2407752. Z.D.L. acknowledges support from STFC studentship ST/Y509346/1. J.R.W. acknowledges that support for this work was provided by The Brinson Foundation through a Brinson Prize Fellowship grant.

\section*{Author Contributions Statement}
D.S.\ led and coordinated the project. C.M.C.\ and J.S.K.\ led the observing proposal. M.F.\ processed the raw JWST observations, and M.S., O.I., H.B.A., J.R.W., and L.P.\ produced the photometric catalogues used in this analysis. D.H.\ measured galaxy shapes. G.L.\ and D.S.\ generated the mass maps using a Kaiser–Squires technique enhanced by S.P.. D.S.\ created the galaxy density map with contribution from A.F.. G.L.\ and D.S.\ identified galaxy clusters. D.S., G.L., D.H., R.M., J.R., and E.H.\ interpreted the maps. D.S., G.L., and R.M.\ wrote the first draft of the paper, on which all authors commented.

\subsection*{Corresponding author}
Correspondence to \href{mailto:dianas@jpl.nasa.gov}{Diana Scognamiglio}.


\section*{Competing Interests Statement}
The authors declare no competing interests.\\

\clearpage

\section*{Supplementary Information} \label{sec:supp_figs}
\begin{figure*}[h!]
      \centering
      \includegraphics[width=\columnwidth]{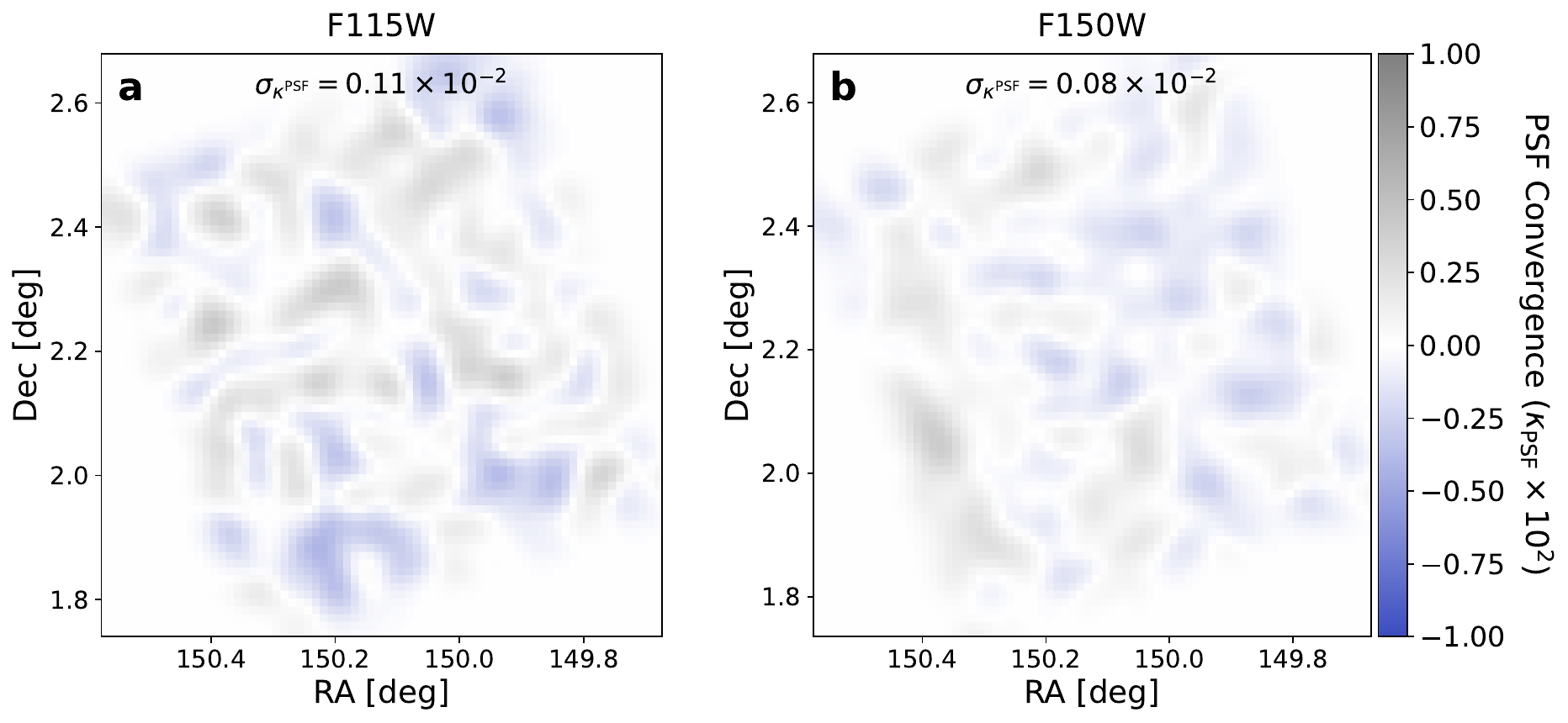} 
      \caption{PSF-induced convergence maps derived from residuals in PSF modelling. \textbf{a}, $E$-mode convergence field $\kappa_{\mathrm{PSF}}$ from F115W-band stellar residuals.  \textbf{b}, Same as in panel a but for the F150W band. The maps display low-amplitude, uncorrelated residuals---over an order of magnitude below the expected lensing signal. The color scale indicates the residual convergence field, illustrating regions where PSF modelling inaccuracies could potentially bias shear measurements.}
    \label{fig:psf_residual_kappa}
\end{figure*}

\begin{figure}[h!]
      \centering
      \includegraphics[width=0.8\columnwidth]{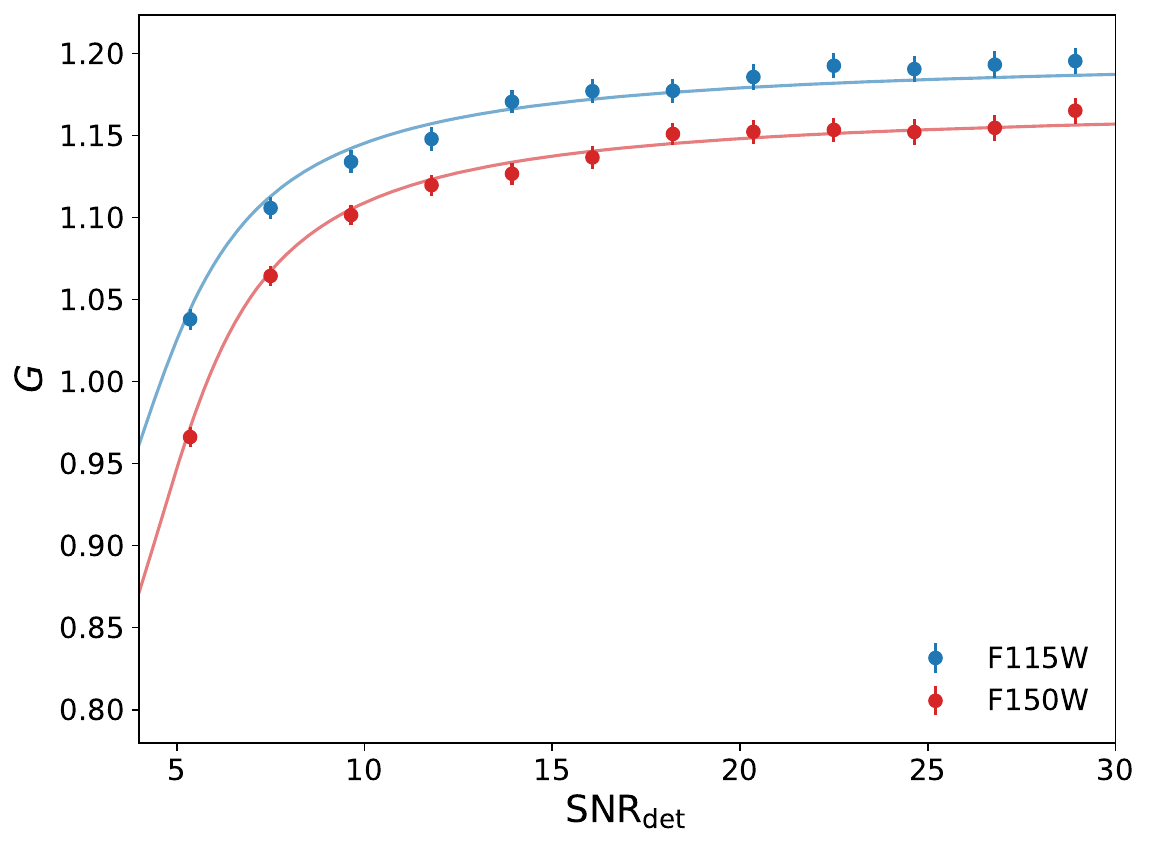} 
      \caption{Signal-to-noise dependence of shear susceptibility of galaxies.
      The shear response $G$ captures the sensitivity of galaxy shape measurements to applied shear and is shown here for F115W and F150W as a function of detection SNR$_{\mathrm{det}}$. Each point shows the mean value of $G$ in bins of SNR$_{\mathrm{det}}$, and the vertical error bars represent $2\sigma$ uncertainties (corresponding to $\sim$95\% confidence).
      The lower asymptotic response in F150W reflects the broader PSF at longer wavelengths.}
\label{fig:shear_susceptibility}
\end{figure}

\begin{figure}[h!]
      \centering
      \includegraphics[width=\columnwidth]{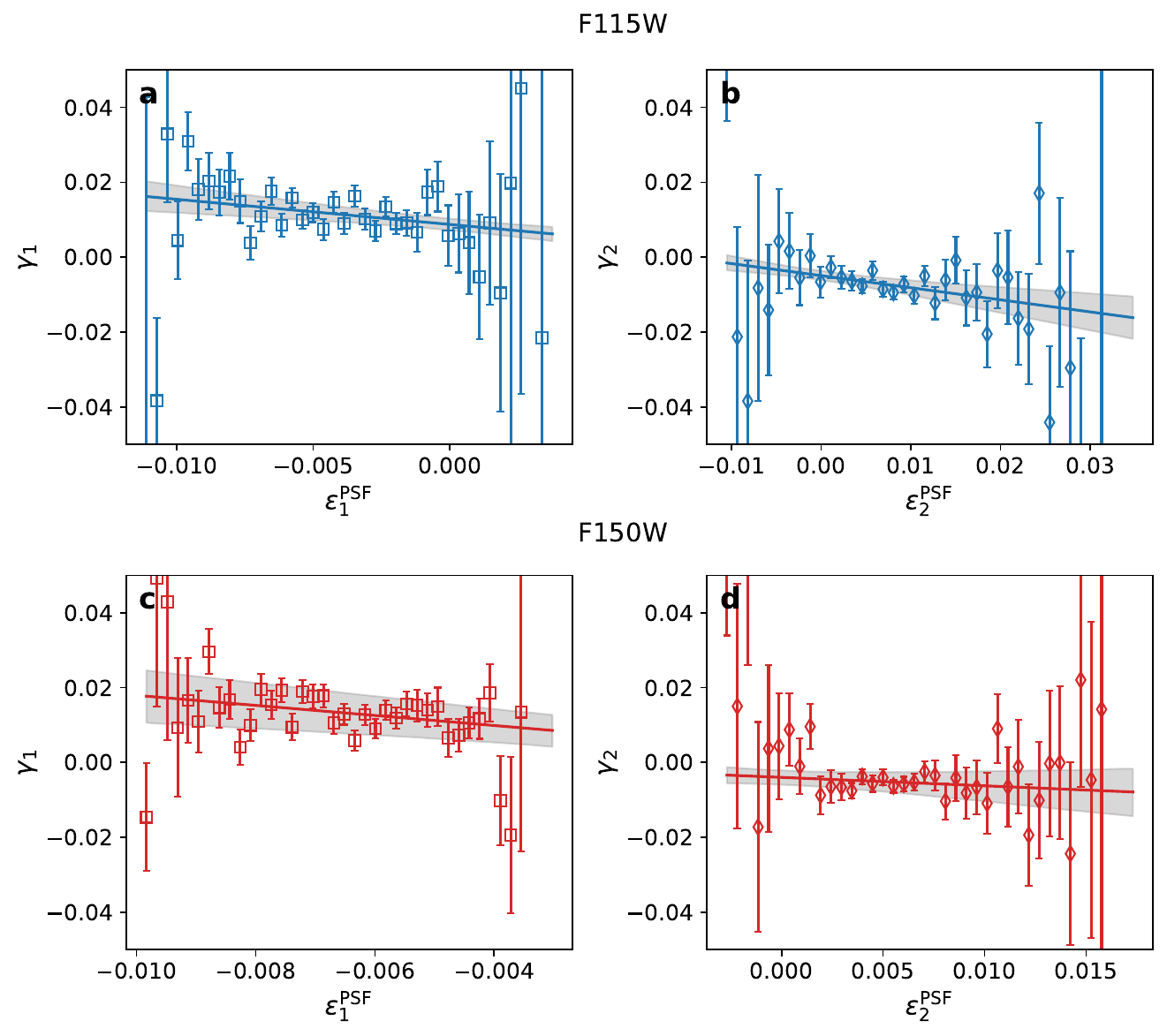} 
      \caption{Residual shear biases as a function of PSF ellipticity in JWST imaging. 
\textbf{a, b}, Average shear values $\gamma_1$ and $\gamma_2$ from galaxies in the F115W band, shown as a function of the PSF ellipticity components $e_1^{\mathrm{PSF}}$ and $e_2^{\mathrm{PSF}}$. 
\textbf{c, d}, The same quantities measured from galaxies in the F150W band. 
Each point represents the mean shear in bins of PSF ellipticity, with vertical error bars showing the standard error of the mean. 
A linear model, $\langle \widehat{\gamma_i} \rangle = a_i + b_i e_i^{\mathrm{PSF}}$, is fitted to the data, with shaded bands indicating 68\% confidence intervals. 
All panels show shear residuals consistent with systematics below the 1\% level.}
\label{fig:psf_measurement_biases}
\end{figure}

\begin{figure}[h!]
      \centering
      \includegraphics[width=\columnwidth]{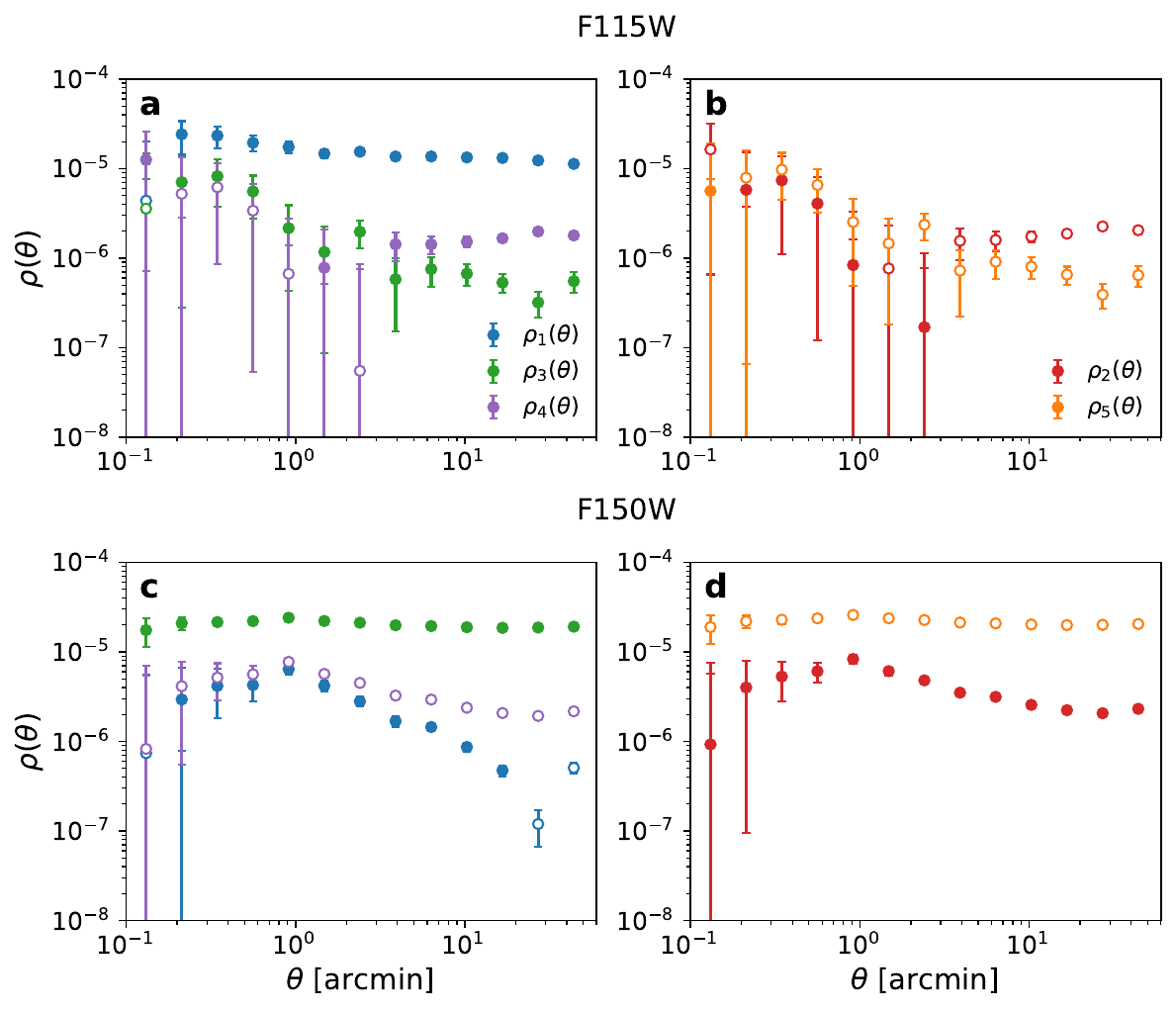} 
      \caption{Residual correlations in PSF-corrected galaxy shapes. \textbf{a}, The $\rho_1(\theta)$, $\rho_3(\theta)$, and $\rho_4(\theta)$ statistics for the F115W band, showing shape correlations due to point-spread function residuals. 
       \textbf{b}, Corresponding $\rho_2(\theta)$ and $\rho_5(\theta)$ for the same band. \textbf{c, d}, The same statistics measured for galaxies in the F150W band. Error bars represent the 1$\sigma$ uncertainty of the $\rho$-statistic in each angular bin. All $\rho_n(\theta)$ functions remain below $10^{-5}$, indicating minimal PSF leakage and high-fidelity shear recovery. Filled and open markers represent measurements in different radial bins, demonstrating the stability of \textsc{WebbPSF} performance across angular scales.}

\label{fig:psf_rho_statistics}
\end{figure}

\begin{table}[ht]
\centering
\captionsetup{justification=justified}
\renewcommand{\arraystretch}{1.4}
\setlength{\tabcolsep}{4pt}
\begin{tabular}{
 l |
 >{\centering\arraybackslash}m{1.5cm} |
 >{\centering\arraybackslash}m{1.5cm} |
 >{\centering\arraybackslash}m{1.1cm} |
 >{\centering\arraybackslash}m{2.7cm} |
 >{\centering\arraybackslash}m{2.7cm} |
 >{\centering\arraybackslash}m{2.7cm}|
 >{\centering\arraybackslash}m{1.5cm}
}
\hline
\textbf{ID}\rule{0pt}{5ex} &
\shortstack{RA \\ {[deg]}}\rule{0pt}{5ex} &
\shortstack{Dec \\ {[deg]}}\rule{0pt}{5ex} &
\raisebox{1.5ex}{\textbf{$z$}}\rule{0pt}{5ex} &
\shortstack{$M_{200} \pm \sigma_{M_{200}}$ \\ {[10$^{13}$ $M_\odot$]}}\rule{0pt}{5ex} &
\shortstack{$R_{200} \pm \sigma_{R_{200}}$ \\ {[deg]}}\rule{0pt}{5ex} &
\shortstack{$L_X \pm \sigma_{L_{X}}$ \\ {[10$^{43}$ erg/s]}}\rule{0pt}{5ex} &
\shortstack{Double \\ Detection}\rule{0pt}{5ex} \\


\hline
20145\text{*} & 149.828 & 2.261 & 0.379  & $2.42 \pm 0.29$ & $0.0279 \pm 0.0011$ & $0.20 \pm 0.038$ & - \\
10220   & 149.921 & 2.522 & 0.729 & $23.3 \pm 0.41$  & $0.0376 \pm 0.00022$ & $11.3 \pm 0.31$ & - \\
30311\text{*} & 149.939 & 2.607 & 0.342 & $3.30 \pm 0.24$  & $0.0339 \pm 0.00082$ & $0.31 \pm 0.037$ & - \\
20193\text{*} & 150.092 & 2.393 & 0.220 & $3.21 \pm 0.15$  & $0.0481 \pm 0.00076$ & $0.26 \pm 0.020$ & - \\

30302 & 150.091 & 2.206 & 0.427 & $1.49 \pm 0.37$ & $0.0217 \pm 0.0018$ & $0.10 \pm 0.042$ & Yes \\
20135 & 150.094 & 2.198 & 0.685 &  $2.89 \pm 0.52$ &  $0.0196 \pm 0.0012$ & $0.41 \pm 0.12$ & Yes \\

20297\text{*} & 150.204 & 2.067 & 0.186 & $1.28 \pm 0.30$  & $0.0405 \pm 0.0031$ & $0.06 \pm 0.029$ & Yes\\
30092\text{*} &  150.211 & 2.070 & 0.370 &  $1.09 \pm 0.35$  & $0.0221  \pm 0.0023 $ & $ 0.06 \pm 0.031 $ & Yes \\

30323  & 150.225 & 2.551 & 1.100 & $5.14 \pm 0.82$  & $0.0175 \pm 0.00091$ & $1.79 \pm 0.46$ & Yes \\
20277 & 150.239 & 2.560 & 0.220 & $2.23\pm0.16 $ & $0.0426\pm 0.0018$ & $0.15 \pm 0.017$ & Yes\\

20153\text{*} & 150.322 & 2.283 & 0.982 & $4.64 \pm 0.68$  & $0.0182 \pm 0.00087$ & $1.28 \pm 0.31$ & - \\
20213 & 150.405 & 2.516 & 0.883 & $4.40 \pm 0.56$  & $0.0187\pm 0.00080$ & $1.03 \pm 0.21$ & - \\

30304 & 150.413 & 2.410 & 0.385 & $5.67 \pm 0.26$  & $0.0371 \pm 0.00055$ & $0.77 \pm 0.06$ & Yes \\
20149 & 150.422 & 2.428 & 0.124 &  $ 6.15 \pm 0.09$ & $0.0986 \pm 0.00046$ & $0.64 \pm 0.014 $ & Yes \\

20127\text{*} & 150.442 & 2.162 & 0.377 &  $1.90 \pm 0.37$ & $0.0262 \pm 0.0016$ & $0.14 \pm 0.040$ & -\\
\hline

\end{tabular}
\vspace{0.5em}
\caption{Galaxy clusters detected with the XMM-\textit{Newton} and Chandra satellites in COSMOS-Web \cite{Gozaliasl2019}. Columns list the cluster ID, right ascension and declination of the X-ray peak (J2000.0), redshift $z$, total mass $M_{200}$ and radius $R_{200}$ with their $\pm1\sigma$ uncertainties (in units of $10^{13}\,M_\odot$ and degrees, respectively), and X-ray luminosity $L_X$ within $r_{500}$ with its $\pm1\sigma$ uncertainty (in units of $10^{43}$ erg\,s$^{-1}$). The last column indicates if the weak lensing detection can be associated with another cluster in the same region. An asterisk indicates galaxy clusters not previously detected in the HST weak lensing map \cite{Massey2007_Nature}.}
\label{tab:new_cluster}
\end{table}

\end{document}